\documentclass[conference]{IEEEtran}
\IEEEoverridecommandlockouts
\usepackage{cite}
\usepackage{amsmath,amssymb,amsfonts}
\usepackage{mathtools}
\usepackage{algorithmic}
\usepackage{graphicx}
\usepackage{textcomp}
\usepackage{bmpsize}
\usepackage{xcolor}
\usepackage{lipsum}
\usepackage{todonotes}
\usepackage{footnote}

\usepackage{multirow}

\usepackage{nth}
\usepackage{tikz}
\usepackage{pgfplots}
\usetikzlibrary{plotmarks}

\usepackage{comment}
\usepackage[utf8]{inputenc}
\usepackage{amsthm}
\usepackage{algorithm2e}

\usepackage{siunitx}

\usepackage{colortbl}
\usepackage{subcaption}

\usepackage{makecell}
\usepackage{booktabs}
\usepackage{float}
\usepackage{csquotes}

\usepackage[misc]{ifsym}
\usepackage[hidelinks]{hyperref}
\def\BibTeX{{\rm B\kern-.05em{\sc i\kern-.025em b}\kern-.08em
    T\kern-.1667em\lower.7ex\hbox{E}\kern-.125emX}}

\usepackage{cleveref}

\usepackage{scrextend}

\newtheorem*{def_adv_ex}{Definition: Adversarial Examples}
    
\begin{document}

\title{Optimizing Information Loss Towards Robust Neural Networks  
    \thanks{Preprint. Under review.}  
}

\author{\IEEEauthorblockN{Philip Sperl}
\IEEEauthorblockA{Fraunhofer AISEC \\
Garching, Germany \\
philip.sperl@aisec.fraunhofer.de}
\and
\IEEEauthorblockN{Konstantin Böttinger}
\IEEEauthorblockA{Fraunhofer AISEC \\
Garching, Germany \\
konstantin.boettinger@aisec.fraunhofer.de}
}

\maketitle

\begin{abstract}
Neural Networks (NNs) are vulnerable to adversarial examples.
Such inputs differ only slightly from their benign counterparts yet provoke misclassifications of the attacked NNs.
The required perturbations to craft the examples are often negligible and even human imperceptible. 
To protect deep learning-based systems from such attacks, several countermeasures have been proposed with adversarial training still being considered the most effective.
Here, NNs are iteratively retrained using adversarial examples forming a computational expensive and time consuming process often leading to a performance decrease.
To overcome the downsides of adversarial training while still providing a high level of security, we present a new training approach we call \textit{entropic retraining}.
Based on an information-theoretic-inspired analysis, entropic retraining mimics the effects of adversarial training without the need of the laborious generation of adversarial examples.
We empirically show that entropic retraining leads to a significant increase in NNs' security and robustness while only relying on the given original data.
With our prototype implementation we validate and show the effectiveness of our approach for various NN architectures and data sets.

\end{abstract}


\section{Introduction}\label{sec:introduction}
Due to the remarkable performance of deep learning, neural networks (NNs) are nowadays used in a wide range of domains.
Their field of operation ranges from software products to consumer electronics and even industrial applications.
Often, NNs are used in security sensitive environments like autonomous driving or intrusion detection systems.
To ensure a secure and safe operation, guarantees on the robustness of the applied NNs are required.
Unfortunately, NNs are known to be vulnerable to so-called adversarial examples \cite{Biggio2013}, \cite{Szegedy2013}, \cite{Nguyen2015a}.
Such slightly perturbed inputs change the classification output of the attacked NNs potentially leading to severe incidents.
These perturbations within adversarial examples are often human imperceptible, which makes the detection and protection against them a difficult task.
In computer vision settings, the change of only one pixel in the input domain often suffices to provoke such misclassifications \cite{DBLP:journals/corr/abs-1710-08864}.
Why adversarial examples exist, and how to reliably defend against them are open questions in the research community.
Recent findings by Ilyas et al. \cite{Ilyas2019} suggest, that there might be a misunderstanding of adversarial examples and how NNs handle features in general.
The authors divide the input space in robust and nonrobust features, whereas the latter are not human understandable and thus exploited by attacks to generate misleading inputs.

Regarding countermeasures, adversarial training is still considered to be the most effective approach. 
To apply this intuitive method, the training set needs to be extended by adversarial examples.
Madry et al. \cite{DBLP:conf/iclr/MadryMSTV18} describe adversarial training as a min-max problem and suggest their PGD attack method as the generation approach during the process.
Yet, additional research is required to further improve adversarial training and explore approaches to overcome its natural downsides.
This includes alternative approaches to the current state of the art of iteratively creating adversarial examples, which is time consuming and has negative effects on the resulting performance of the hardened NNs.
An important contribution marking the first step into this direction was presented by Shafahi et al. \cite{Shafahi2019}.
The authors introduce the first adversarial-example-free training procedure, achieving high levels of robustness.
In a similar spirit, we present a new information-theoretic-inspired approach.

In this paper, we first shed light on adversarial training and analyze its impacts on NNs.
Here, we profit from two intriguing research directions which gained attention in recent years.
On the one hand, we build upon the findings in activation analysis.
Recent work in this area analyzed the hidden activation values of NNs to gain insights about the processed inputs \cite{Sperl2019}, \cite{SperlSchulze2020}.
It is shown that the hidden activations carry information highly relevant for a security-related analysis of NNs.
Additionally, we take into account approaches using information-theoretic concepts to understand NNs \cite{Tishby2015}, \cite{Shwartz-Ziv2017}, \cite{Gabrie2018}, \cite{Amirian}, \cite{Terzi2020}. 
Combining these two worlds reveals new aspects of how neural networks behave under different circumstances and types of inputs.
This fusion leads to a new training approach, which we call \textit{entropic retraining}.
We empirically show, that this training procedure mimics the effects of adversarial training, significantly improving the robustness of NNs.
Interestingly, this approach does not require adversarial examples and works for various NN architectures and data sets using the original input data only.

In summary we make the following contributions:
\begin{itemize} \setlength\itemsep{-0.2em}
    \item We analyze the impacts of adversarial training on NNs using information-theoretic-inspired approaches.
    \item We leverage the knowledge of this analysis and present a new training method that we call \textit{entropic retraining}.
    \item We implement and evaluate our new training method and show a significant improvement in robustness for various NN architectures and two data sets.
\end{itemize}

\section{Background and Related Work} \label{related_work}
\subsection{Adversarial Examples} \label{adversarial_attacks}
In this paper, we introduce a new training approach for NNs mimicking the effects of adversarial training and thus significantly increasing robustness.
This level of robustness describes the potential of an attacker performing so-called evasion attacks.
Here, during test-time only the inputs to the attacked NNs may be changed.
The aim of such an evasion attack is to generate adversarial examples which are close to their benign counterparts but still provoke misclassifications. 
Formally, adversarial examples can be defined as follows:
\begin{def_adv_ex}
Let f($\cdot$) be a trained NN performing classifications. 
Let H($\cdot$) be a human oracle performing the same classifications and similar capabilities such that for a given benign input $\textbf{x}$ the following holds:
\begin{equation}
f(\textbf{x}) = H(\textbf{x})
\label{eq1}
\end{equation}
Let $\textbf{x'}$ be a permuted version of $\textbf{x}$ under the constraint of visual similarity, i.e., $\|\textbf{x'}-\textbf{x}\| \leqslant \epsilon$ for some small $\epsilon\in\mathbb{R^+}$. 
Then $\textbf{x'}$ is an adversarial example, if the following holds:
\begin{equation}
H(\textbf{x}) = H(\textbf{x'})\  \  \wedge \  f(\textbf{x'}) \neq H(\textbf{x'}).
\end{equation}
\label{eq2}
\end{def_adv_ex} 

Test-time evasion attacks can be divided into white-box and black-box approaches which differ in the amount of information available for the attacker \cite{DBLP:journals/corr/abs-1810-00069}.
Whereas in white-box attacks the attacker has complete access to the attacked model, in black-box attacks only the input-output relation is visible.
In this paper, we focus on the more effective white-box attacks such as the ``Fast Gradient Sign Method'' (\textit{FGSM}) \cite{43405}, or the ``Projected Gradient Descent'' (\textit{PGD}) \cite{DBLP:conf/iclr/MadryMSTV18} attack.
The currently most effective white-box approach, the \textit{C$\&$W} attack, was introduced by Carlini and Wagner \cite{DBLP:journals/corr/CarliniW16a}.
Here, a cost function $f_{y}$ is introduced as an optimization substitute opposed to previous approaches were the loss function was used directly.
In \Cref{sec:experiments_attack_methods} we introduce the attacks we considered throughout this paper in more detail.

\subsection{Robustness Quantification}\label{sec:how_robust}
State-of-the-art research on NN robustness still lacks meaningful and generally applicable robustness metrics.
Currently, $l_{p}$-norms are used to calculate the distance between adversarial examples and their benign counterparts as an indicator for the robustness of the analyzed NN.
It is assumed that high $l_{p}$-based distances directly correlate with the exhibited level of robustness.
In the majority of cases, this approach poses a viable solution, yet it is shown that new robustness metrics are needed to cover a wider variety of possible scenarios \cite{Sharif2018}, \cite{Sen2019}.
Recent work has provided some alternatives, for example Wang et al. with their CleverScore \cite{Weng2018c}.
While being a viable solution quantifying the robustness of a wide range of NNs, this metric cannot be applied to all architectures.
Therefore, in this paper, we stick to the traditional approach of measuring NNs' robustness using $l_{p}$-norms.
More precisely, we craft adversarial examples with constant $l_{p}$-distances compared to their benign counterparts.
Subsequently, we measure the fraction of successful adversarial examples, i.e. examples which lead to misclassification, and refer to this fraction as the attack success rate in the following.
This approach is still widely used in adversarial machine learning and thus allows a direct comparison to related research.

\subsection{Adversarial Training}
Adversarial training is still considered to be the most effective countermeasure to protect NNs against evasion attacks.
Even though adversarial training is well researched and evaluated, some aspects still remain unclear: for instance, the full impact on the protected NNs and how to further improve the concept of adversarial training in general.
The trade-off between the resulting level of robustness and final accuracy of the model is known, but yet to be fully understood.
Furthermore, the adequate setting of hyperparameters poses an uncertainty for developers trying to protect their NNs using adversarial training.
Here, the number of epochs, learning rate, or attack parameters need to be defined and are often set following a laborious empirical approach.
Recent work by Shafahi et al. \cite{Shafahi2019} showed that adversarial training can be further improved and even performed at reduced cost.
The authors reuse the gradient information available during the updates of the model parameters during normal training.
With this approach, the authors achieve a similar level of robustness compared to PGD-based adversarial training, without the generation of adversarial examples.

Still, to shed more light on adversarial training, we inspect NNs using information-theoretic-inspired approaches at different robustness levels.
In this paper, we follow the adversarial training approach presented by Madry et al. using the PGD attack method \cite{DBLP:conf/iclr/MadryMSTV18}.
During this iterative process the following min-max problem is solved:
\begin{equation}
\min_{\theta} \rho(\theta).
\end{equation}
with 
\begin{equation}
\rho(\theta) = \mathbb{E}_{(x,y) \sim D}[\max_{\delta \in S} L(\theta, x + \delta,y)]
\end{equation}

\subsection{Activation Analysis}
In this new research direction, the hidden activation values of NNs are analyzed to observe their behavior and detect irregularities in the input space.
Inspired by neuron coverage guided testing \cite{DBLP:journals/corr/PeiCYJ17, DBLP:journals/corr/abs-1803-07519, DBLP:journals/corr/abs-1803-04792,tensorfuzz} several new methods have been proposed, directly analyzing the hidden activations.
For example Sperl et al. \cite{Sperl2019} observe the hidden activations of NNs while processing benign and adversarial inputs.
By training a secondary NN on the extracted hidden activations acting as a binary classifier, the authors are able to detect adversarial examples fed to the initial NN.
Using their method, the authors are able to reliably detect attacks for various NN architectures, data sets, and types of data.
Further refined and improved, Sperl and Schulze \cite{SperlSchulze2020} use this approach in the more general semi-supervised setting of anomaly detection.
The authors are able to detect anomalies even with a small amount of anomalous samples available during training.
These findings indicate that the hidden activations of NNs carry security-sensitive information highly relevant for the understanding of the robustness of such systems.

Throughout this paper, we take into account previous research in this field and therefore introduce the concept in a more formal manner.
As presented in \Cref{adversarial_attacks}, NNs can be described by $f(\textbf{x}; \textbf{$\theta$}) = \hat{\textbf{y}}$, where $\textbf{$\theta$}$ are the learned parameters, $\textbf{x}$ the inputs, and $\hat{\textbf{y}}$ the predictions of the NN.
With respect to the layers, this function can be written as $f=f_1 \circ \ldots \circ f_M$.
For each layer, $f_i(\textbf{x}; \textbf{$\theta$}) = \textbf{h}_i$ describes the layer output after the application of the activation function.
We call the concatenation of all outputs as the activation values based on which we present our new training approach.

\subsection{Related Work}
In the following we shortly present the most related publications using information-theoretic approaches to study NNs.
We focus on work improving the understanding of NNs in general and especially the robustness against adversarial examples.
In 2000, Tishby et al. \cite{Tishby2000} introduced the information bottleneck (IB) method solving the problem of distributional clustering.
The findings were further refined, e.g. by Tishby et al. \cite{Tishby2015}, and used to shed light on the training of NNs.
Shwartz-Ziv and Tishby \cite{Shwartz-Ziv2017} introduced their so-called information planes as a metric describing the information flow during NN training.
The authors analyze the mutual information between the layers and make the following observations:
First, NNs perform a compression of the inputs to a compact representation.
Then, by neglecting or \textit{forgetting} parts of the compressed information, NNs achieve better generalization and overall performance.
In 2019, Achille et al. \cite{Achille2019} investigated NNs using information-theoretic approaches as well and make similar observations.
The authors show that models which generalize well, often show a low level of information.

Based on these findings and concepts, we investigate the impacts of adversarial training on NNs.
Closely related in this regard are the contributions by Terzi et al. \cite{Terzi2020} published in 2020, showing that adversarial training reduces information in NNs.
Using this insight, the authors empirically show that robust models show a better transferability characteristic compared to their unsecured counterparts for the CIFAR-10 and CIFAR-100 data sets.
In this paper, we profit from this finding and present a new training approach we call entropic retraining.

\section{Threat Model}\label{threat_models}
Following the guideline on the evaluation of adversarial defense methods by Carlini et al. \cite{Carlini2019} we present the threat model we considered throughout our experiments.
We elaborate on the conditions under which we tested the robustness of our countermeasure to give the reader a better insight and provide the basis for future work.
In the case of adversarial machine learning, a threat model describes the following aspects and attacker capabilities: 
\begin{itemize}
    \item Goals of the adversary
    \item Capabilities of the adversary
    \item Knowledge of the adversary
\end{itemize}

The goal of an attacker can either be to alter the classification output in a non-targeted manner to an arbitrary class or to lead the NN to predict a chosen, targeted class.
Here, we solely evaluate non-targeted attacks.
We expect that the experiments as well as the results can be directly transferred to the non-targeted setting.

When performing evasion attacks, the capabilities of the adversary are limited to changes in the input space during run-time.
Hence, the network parameters and weights are not accessible and remain unchanged during the attack process.
The impact of the changes on the input performed by an attacker are usually measured using $l_{p}$-norms as described earlier.
In image classification tasks and machine learning in general, typically the $l_{0}$, $l_{2}$, or $l_{\infty}$ distances are considered.
In this paper, we solely utilize attack methods which are bound by the $l_{2}$ and $l_{\infty}$ norm.

Finally, the adversary's knowledge can range from a black-box setting where solely the input-output behaviour of the model is visible, to a fully white-box setting.
In this case, the attacker additionally observes the model's gradients, activations, and weights to specifically craft attacks for the architecture leveraging all available information.
We assume attackers with white-box capabilities.
Hence, we use specifically crafted adversarial examples to lead our models under attack to misclassifications.
Carlini and Wagner \cite{Carlini:2017:AEE:3128572.3140444} suggested that when presenting new defense strategies, the robustness against \textit{adaptive attacks} needs to be evaluated.
In this attack setting, adversaries adapt their attack methods to exploit the characteristics of the applied defense strategies.
Even though we assume the attackers to be aware of our modified training method, further attack methods targeting the information-theoretic properties of the models under attack are out of scope of this paper and left for future work.

\section{Entropic Retraining} \label{method}
In this section we introduce our main contribution, a modified training approach we call entropic retraining.
Entropic retraining mimics the effects of adversarial training, providing a comparable increase in the robustness of NNs.
Based on the adapted loss function and optimization goal, entropic retraining does not rely on the time consuming and computational expensive process of generating adversarial examples.

For better understanding, we first evaluate adversarial training and the behavior of the resulting robust NNs.
We then transfer our findings to an adversarial example free case.

\subsection{Preliminary Analysis}
We assess the impact of adversarial and entropic training using two information-theoretic-inspired values.
Mainly, we were inspired by the classical Shannon entropy \cite{Shannon1948} defined as:
\begin{equation}
    H(x) = - \sum_{i}^{} p(x_{i})\log(p(x_{i})).
\end{equation}
Here, we directly use the hidden activation values of each layer of the NN.
More precisely, we extract the activation values $\textbf{h}_i$ of each layer $i \in {1,..,M}$ and calculate $E(\textbf{h}_i)$ as follows:
\begin{equation}
    E(\textbf{h}_i) = - \sum_{j=1}^{N_i} h_{j}\log(h_{j}),
\end{equation}
with $N_i$ being the number of neurons in layer $i$.
To finally calculate the network-entropy-related value $E_{N}$ we first normalize $E(\textbf{h}_i)$ with respect to the number of neurons in the layer and then average the result with respect to the number of layers in the NN:
\begin{equation}
    E_{N} = \dfrac{1}{M} \sum_{i=1}^{M} \dfrac{1}{N_{i}} E(\textbf{h}_i).
\end{equation}

Moreover, we introduce and evaluate $E_{T}$, which layer-wise quantifies the differences in $E(\textbf{h}_i)$ and thus allows an evaluation of the information-related flow through the NN.
More formally, we define $E_{T}$ as follows:
\begin{equation}
\begin{aligned}
    E_{T} = \dfrac{1}{M-1} \sum_{i=1}^{M-1} E(\textbf{h}_i) - E(\textbf{h}_{i+1}) \\
    = \dfrac{1}{M-1} (E(\textbf{h}_1) - E(\textbf{h}_{M}))
\end{aligned}
\end{equation}

Note, that we do not normalize the $E(\textbf{h}_i)$ values to calculate $E_{T}$.
We argue that non-normalized values provide a more useful picture of the information-related flow in this case.
During the adversarial training process and the solving of the earlier introduced min-max problem some of the weights might be set to negligible magnitudes.
This would lead to neurons with an insignificant impact on the resulting level of information.
Hence, normalizing $E(\textbf{h}_i)$ would hide this effect.
We base our assumption on the contribution by Ilyas et al. \cite{Ilyas2019}.
The authors argue, that models require to \textit{forget} some of the non-robust features to be more robust.
This correlates to neglecting information of certain neurons trained to process these non-robust features.
To capture this effect and the overall impact of adversarial training on the NNs, we do not normalize the components of $E_{T}$.

In summary, with our newly introduced values $E_{N}$ and $E_{T}$ at hand we evaluate the information-theoretic-related properties of the NNs during input processing.

\subsection{Impacts of Adversarial Training}\label{sec:impact_adv_training}
For our experiment in which we want to visualize the impact of adversarial training, we considered the following settings and model under test (MUT).
We used a pretrained Lenet 5 CNN \cite{LeCun:1989:BAH:1351079.1351090} model.  
This architecture is designed to classify the simple and widely used MNIST data set.
For the sake of readability, we introduce the architecture and data set in more detail in \Cref{sec:experiments_nns}.

\begin{figure}[]
  \begin{center}
  \begin{tikzpicture}

% removing exponent from plot

\begin{axis}[
yticklabel style={
        /pgf/number format/fixed,
        /pgf/number format/precision=5
},
ylabel style={rotate=-90},
scaled y ticks=false,
xlabel={Number of Epochs in Adversarial Training},
ylabel={$E_{N}$}
]
\addplot+[black, mark options={fill=black}] table[x=Num Advs, y=B_DLA_NE_mean, col sep=comma] {data/MNIST_00_lenet_5_LinfFastGradientAttack_LinfPGD_results.csv};

\end{axis}
\end{tikzpicture}
    \caption{$E_N$ for the Lenet 5 Model classifying the MNIST data set during adversarial training.}
    \label{fig:adv_train_entropy}
  \end{center}
\end{figure}
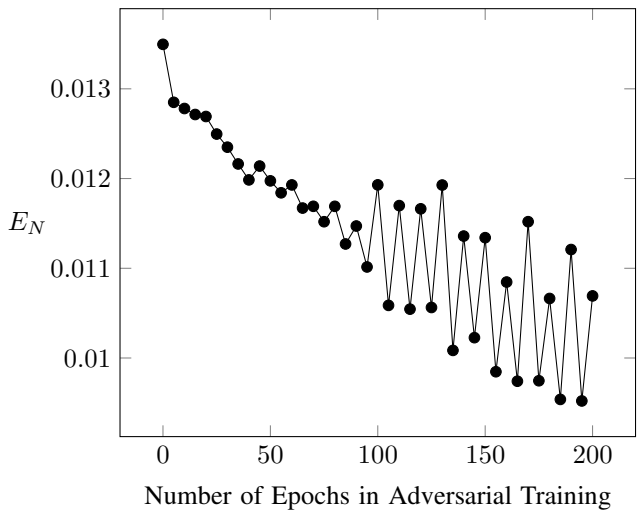

\begin{figure}[]
  \begin{center}
  \begin{tikzpicture}
\begin{axis}[
ylabel style={rotate=-90},
xlabel={Number of Epochs in Adversarial Training},
ylabel={$E_{T}$}
]
\addplot+[black, mark options={fill=black}] table [x=Num Advs, y=B_DLA_ET_mean, col sep=comma] {data/MNIST_00_lenet_5_LinfFastGradientAttack_LinfPGD_results.csv};

\end{axis}
\end{tikzpicture}
    \caption{$E_T$ for the Lenet 5 Model classifying the MNIST data set during adversarial training.}
    \label{fig:adv_train_entropy_transition}
  \end{center}
\end{figure}
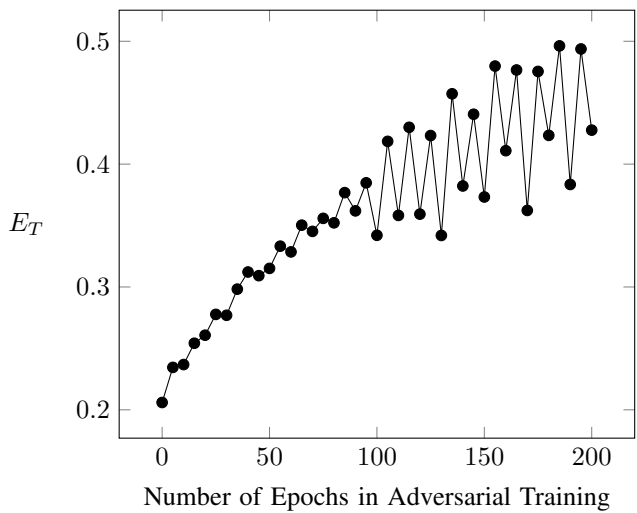

During the adversarial training process, we calculated $E_{N}$ and $E_{T}$ after every five epochs.
In summary, we adversarially trained the MUT for $200$ epochs.
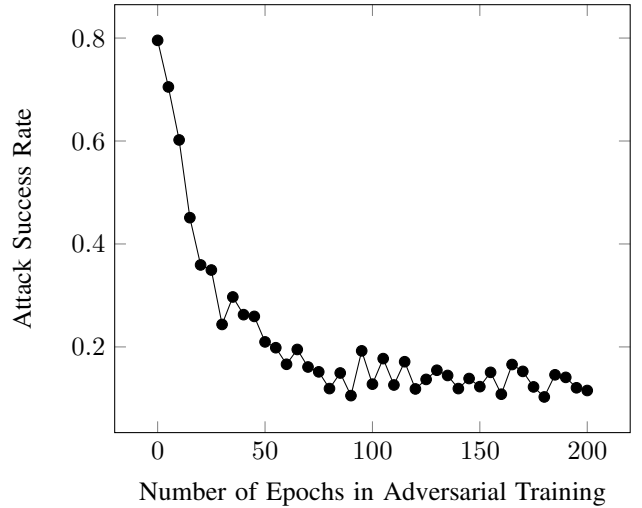
\begin{figure}[]
  \begin{center}
  \begin{tikzpicture}
\begin{axis}[
xlabel={Number of Epochs in Adversarial Training},
ylabel={Attack Success Rate}
]
\addplot+[black, mark options={fill=black}] table [x=Num Advs, y=adv_success, col sep=comma] {data/MNIST_00_lenet_5_LinfFastGradientAttack_LinfPGD_results.csv};

\end{axis}
\end{tikzpicture}
    \caption{Evolution of the robustness during adversarial training of the Lenet 5 model classifying MNIST when attacked using FGSM.}
    \label{fig:adv_train_attack_success}
  \end{center}
\end{figure}
In \Cref{fig:adv_train_attack_success} we show the resulting attack success rate for each instance of the MUT.
To quantify the model's level of robustness, we generated $l_{\infty}$-bound ($\epsilon$ = $0.3$) adversarial examples using FGSM with images randomly drawn from the test set.
Due to this limitation in the permutation space, not all images resulted in a successful misclassification, allowing the calculation of the attack success rate.
As expected, for a higher number of epochs in adversarial training we see a decrease in the attack success rate, indicating a robustification of the MUT.
With Figures \ref{fig:adv_train_entropy} and \ref{fig:adv_train_entropy_transition} we visualize the impact of adversarial training on the same instances of the MUT using the previously introduced $E_N$ and $E_T$.
In \Cref{fig:adv_train_entropy} we see a decrease in $E_N$ allowing the following conclusion which strongly corresponds to the contribution of Terzi et al. \cite{Terzi2020}: 
\begin{displayquote}
\textit{Adversarial training decreases $E_N$.}
\end{displayquote}
This effect is already visible after five epochs of adversarial training.
Furthermore, a strong correlation between the evolution of the attack success rate and $E_N$ is given.
As stated before and corresponding to previous work \cite{Ilyas2019, Terzi2020}, adversarial training increases the model's robustness by boosting the generality of the MUT at the cost of its initial accuracy \cite{Zhang2019b}.

In \Cref{fig:adv_train_entropy_transition} we observe a similar behavior of the NNs when observing $E_T$.
The plot shows, that $E_T$ increases with higher level of robustness. 
Together with the previous observation of decreasing attack success rates, this suggests a negative correlation between $E_T$ and the level of robustness achieved during adversarial training.
This effect is even more visible for robust models, which brings us to the following assumption:
\begin{displayquote}
\textit{Adversarial training increases $E_T$.}
\end{displayquote}

\subsection{Entropic Retraining Objective}
With the above figures and observations at hand, we present the concept of entropic retraining.
Our new training concept mimics the effects of adversarial training, without the need of adversarial examples.
Above, we observed that adversarial training yields NNs with lower levels of $E_N$ and higher levels of $E_T$.
This motivates us to incorporate $E_T$ in an adapted loss function.
During entropic retraining, we optimize the performance of the model while simultaneously increasing $E_T$.
Formally, we define our adapted loss function as follows:
\begin{equation}
    L_{tot} =
    L_{scce}(\textbf{y}, \hat{\textbf{y}})  + 
    \lambda \cdot L_{T}(E_{T}, \hat{E}_{T}).
\end{equation}
with 
\begin{equation}
    L_{scce}(\textbf{y}, \hat{\textbf{y}}) = - \sum_{i=1}^{N} y_{i} log(p_{\hat{y_{i}}}).
\end{equation}
being the standard sparse categorical cross-entropy and
\begin{equation}
    L_{T}(E_{T}, \hat{E}_{T}) = {(\hat{E}_{T} - \sigma E_{T})}^2.
\end{equation}

Here, $E_{T}$ is the value of the original pretrained model, while $\hat{E}_{T}$ is calculated during entropic retraining.
The factors $\lambda,\sigma \in \mathbb{R}^+$ are positive hyperparameters set in the optimization process before training.

\subsection{Impacts of Entropic Retraining}
\begin{figure}[]
  \begin{center}
  \begin{tikzpicture}
\begin{axis}[
yticklabel style={
        /pgf/number format/fixed,
        /pgf/number format/precision=5
},
ylabel style={rotate=-90},
scaled y ticks=false,
xlabel={Number of Epochs in Entropic Retraining},
ylabel={$E_{N}$}
]
\addplot+[red, mark options={fill=red}] table [x=Num Advs, y=B_DLA_NE_mean, col sep=comma] {data/MNIST_00_lenet_5_LinfFastGradientAttack_DLA_results.csv};

\end{axis}
\end{tikzpicture}
    \caption{$E_N$ for the Lenet 5 Model classifying the MNIST data set during entropic retraining.}
    \label{fig:dla_train_entropy}
  \end{center}
\end{figure}
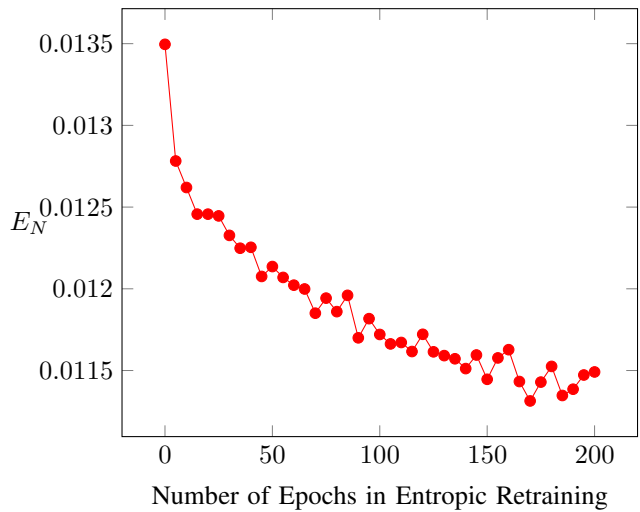

\begin{figure}[]
  \begin{center}
  \begin{tikzpicture}
\begin{axis}[
ylabel style={rotate=-90},
xlabel={Number of Epochs in Entropic Retraining},
ylabel={$E_{T}$}
]
\addplot+[red, mark options={fill=red}] table [x=Num Advs, y=B_DLA_ET_mean, col sep=comma] {data/MNIST_00_lenet_5_LinfFastGradientAttack_DLA_results.csv};

\end{axis}
\end{tikzpicture}
    \caption{$E_T$ for the Lenet 5 Model classifying the MNIST data set during entropic retraining.}
    \label{fig:dla_train_entropy_transition}
  \end{center}
\end{figure}
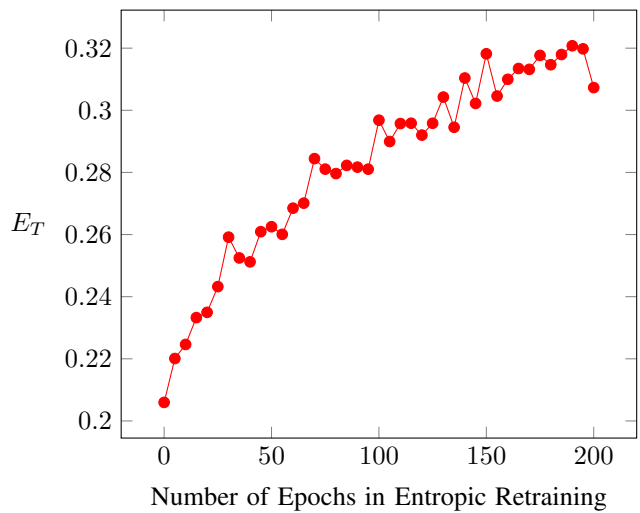
In this section we show the impacts of entropic retraining with the following experiment:
We performed entropic retraining with the same pretrained MUT from above based on the Lenet 5 architecture.
During this process, we calculated $E_N$ and $E_T$.
To assess the achieved level of robustness after every five epochs, we attacked the resulting instances of the MUT and measured the achieved attack success rates.
In Figures \ref{fig:dla_train_entropy} and \ref{fig:dla_train_entropy_transition} we show the evolution of $E_N$ and $E_T$, respectively.
Compared to adversarial training, we see the same effects for the resulting robust model.
With our experiments we make the following observation:
\begin{displayquote}
\textit{Entropic retraining decreases $E_N$.}
\end{displayquote}
Moreover:
\begin{displayquote}
\textit{Entropic retraining increases $E_T$.}
\end{displayquote}
This observation suggests that entropic retraining indeed mimics adversarial training with respect to the introduced values $E_N$ and $E_T$.
Remarkably, we achieve this only using the original inputs without the generation of adversarial examples.

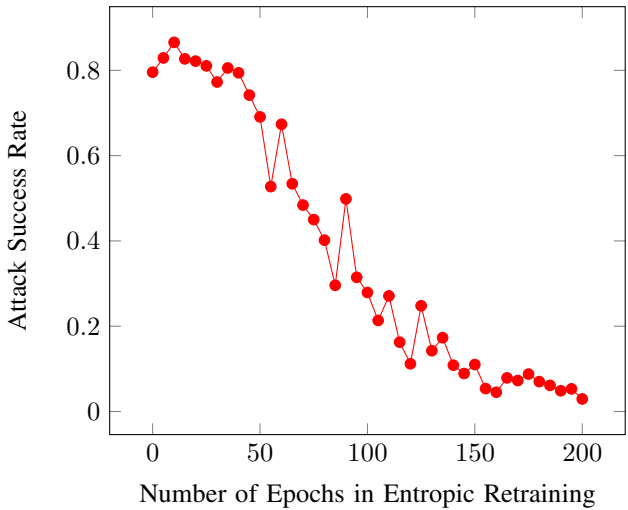
\begin{figure}[]
  \begin{center}
  \begin{tikzpicture}
\begin{axis}[
xlabel={Number of Epochs in Entropic Retraining},
ylabel={Attack Success Rate}
]
\addplot+[red, mark options={fill=red}] table [x=Num Advs, y=adv_success, col sep=comma] {data/MNIST_00_lenet_5_LinfFastGradientAttack_DLA_results.csv};

\end{axis}
\end{tikzpicture}
    \caption{Evolution of the robustness during entropic retraining of the Lenet 5 model classifying MNIST when attacked using FGSM.}
    \label{fig:dla_train_attack_success}
  \end{center}
\end{figure}

In \Cref{fig:dla_train_attack_success} we show the attack success rates for the model after entropic retraining.
Again, the same effect as in the previously performed adversarial training is visible.
The attack success rate decreases with the number of performed entropic retraining epochs.
With this observation, we provide first strong indicators, that an analysis of the information-theoretic behavior of NNs gives insights about their level of robustness.

\section{Implementation and Experimental Setup} \label{Experiments}
In the following we extend our initial observation to a thorough and statistically sound evaluation.
Hence, we introduce details on our implementation, considered data sets, and evaluated NN architectures.
For our prototype implementation we used Keras \cite{chollet2015keras} and ran the training procedures on an Intel Xeon E5-2640 v4 server with an NVIDIA Titan X GPU.

\subsection{Considered Data Sets}
We used the MNIST \cite{Lecun1998a} and CIFAR-10 \cite{krizhevsky2009learning} image datasets.
The two data sets are commonly used in machine learning research and allow an easy comparison to existing countermeasures, especially adversarial training.
MNIST consists of $70000$ handwritten digits ranging from $0$ to $9$ represented by $28\times28$ gray-scale pixels.
CIFAR-10 consists of $60000$ colored images represented by $32\times32\times3$ pixels.

\subsection{Considered Neural Networks}\label{sec:experiments_nns}
Throughout our evaluation we considered three different NN architectures.
For the sake of comparability and to ease reproducibility we utilized publicly available and thoroughly evaluated architectures.
For MNIST, we considered the Lenet 5 architecture \cite{LeCun:1989:BAH:1351079.1351090} and an example model provided by Keras \cite{kerasEx_MNIST}.
Similarly, for CIFAR-10 we used the example model provided by Keras \cite{kerasEx_CIFAR10}.
\begin{table*}[tb]
\centering
\caption{\sc{Architectures and Performance of the Considered Neural Networks}}
\begin{tabular}{ll>{}m{4cm}>{}m{3cm}l}
\toprule
\textbf{Data Set}         & \textbf{Model} & \textbf{Model Details} & \textbf{Training Settings} & \textbf{Test Accuracy} \\ \hline
\multirow{4}{*}{MNIST}   & Lenet 5 \cite{LeCun:1989:BAH:1351079.1351090}              & -- 2 convolutional layers with filter sizes 5 and 16 \newline -- each convolutional layer is followed by a average-pooling layer \newline -- finally, 3 dense layers with 120, 84, and 10 neurons each \newline -- no drop-out layers    & -- optimizer: adam \newline -- learning rate: 0.001 \newline -- epochs = 24 \newline -- batch size: 64                   & 99.05$\%$                  \\ \cline{2-5}
                         & MNIST Keras CNN \cite{kerasEx_MNIST} & -- 2 convolutional layers with filter sizes 32 and 64 \newline -- followed by 1 max-pooling layer \newline -- finally, 3 dense layers with 512, 84, and 10 neurons each \newline -- no drop-out layers
& -- optimizer: adam \newline -- learning rate: 0.001 \newline -- epochs = 12 \newline -- batch size: 128 & 99.13$\%$                  \\  \hline
\multirow{1}{*}{CIFAR-10} & CIFAR Keras CNN \cite{kerasEx_CIFAR10} &    
 -- 4 convolutional layers, the first two with filter of size 32, the second pair with filter size of 64 \newline -- each pair of convolutional layers is followed by a max-pooling layer \newline -- finally, 6 dense layers with 512, 256, 128, 128, 84, and 10 neurons each \newline
 -- no drop-out layers
 & -- optimizer: RMSprop \newline -- learning rate: 0.0001 \newline -- epochs = 50 \newline -- batch size: 32 \newline -- using data augmentation                      & 83.10$\%$                  \\ \bottomrule 
\end{tabular}
\label{target_models}
\end{table*}
In Table \ref{target_models} we summarize the considered NNs and show details regarding their architectures and the executed training processes.
Note, that for both Keras examples models, we slightly changed the architectures and neglected the drop-out layers.

\subsection{Considered Attack Methods}\label{sec:experiments_attack_methods}
We considered the following four attack methods: PGD\cite{DBLP:conf/iclr/MadryMSTV18}, BIM\cite{DBLP:journals/corr/KurakinGB16}, FGSM\cite{43405}, and C$\&$W\cite{DBLP:journals/corr/CarliniW16a}.
For the PGD attack we considered both, the $l_{2}$ and $l_{\infty}$ bound versions which we call L2PGD and LinfPGD in the following.
BIM and FGSM are $l_{\infty}$ bound while we chose the $l_{2}$ bound version of the C$\&$W attack.
We motivate our choice of attacks based on their performance and popularity and to provide a diverse assessment of the quality of our proposed countermeasure.
Note, that we used the \textit{foolbox} framework \cite{rauber2017foolbox} to generate adversarial examples using the above stated methods.
For MNIST we limit $\epsilon$ to $4.5$ and $0.3$ for $l_{2}$ and $l_{\infty}$ bound attacks, respectively.
Similarly, for CIFAR-10 we limit $\epsilon$ to $0.95$ and $0.03$ for $l_{2}$ and $l_{\infty}$ bound attacks, respectively.

\subsection{Entropic Retraining Settings}
In \Cref{target_models_lossy_training} we summarize the settings we chose to perform entropic retraining for the three MUTs.
\begin{table}[tb]
\centering
\caption{\sc{Entropic Retraining Settings for the Considered Neural Networks}}
\begin{tabular}{ll>{}m{3.5cm}}
\toprule
\textbf{Data Set}         & \textbf{Model} & \textbf{Entropic Retraining Settings} \\ \hline
\multirow{4}{*}{MNIST}   & Lenet 5 & -- optimizer: adam \newline -- learning rate: $0.0009$ \newline -- epochs = $200$ \newline -- batch size: $4$ \newline -- $\lambda$: $1$ \newline -- $\sigma$: $1.65$                      \\ \cline{2-3}
                         & MNIST Keras CNN 
& -- optimizer: adam \newline -- learning rate: $0.001$ \newline -- epochs = $200$ \newline -- batch size: $4$ \newline -- $\lambda$: $1$ \newline -- $\sigma$: $1.65$                      \\  \hline
\multirow{1}{*}{CIFAR-10} & CIFAR Keras CNN  & -- optimizer: adam \newline -- learning rate: $0.0005$ \newline -- epochs = $500$ \newline -- batch size: $64$ \newline -- $\lambda$: $1$ \newline -- $\sigma$: $1.01$ \\ \bottomrule 
\end{tabular}
\label{target_models_lossy_training}
\end{table}
Setting the parameters, we followed an intuitive approach and did not perform an extensive search.
Hereby, we want to show the simplicity of our approach underlining the ease of use.
We used the adam optimizer \cite{kingma2014adam} to train the three models and chose similar settings in each case.
As the gradient calculation during entropic retraining poses a rather complex task, we used small batches.
With this, we aim to avoid sharper minimizers during training, which would lead to a decrease in generalization as shown by Keskar et al. \cite{Keskar2019}.
For both MNIST models we chose a batch size of four, while the CIFAR-10 model was trained with a batch size of 64.
Accordingly, we set the number of epochs to 200 and 500 for the MNIST and CIFAR-10 experiments, respectively.
Finally, we set the weighting factor $\sigma$ based on the observations shown in \Cref{sec:impact_adv_training}.
As we want to increase $E_T$, we chose $\sigma>1$.

\begin{table*}[h!]
\centering
\caption{\sc{Attack success rates when attacking the original neural networks}}
\begin{tabular}{lllllll}
\toprule
\multirow{2}{*}{\textbf{Data Set}} & \multirow{2}{*}{\textbf{Model}} & \multicolumn{5}{l}{\textbf{Attack Success Rates using the following methods:}}                                  \\ \cline{3-7} 
                                  &                                        & \textit{\textbf{LinfPGD}} & \textit{\textbf{BIM}} &  \textit{\textbf{FGSM}} & \textit{\textbf{C\&W}}\footref{CW_footnote}  &  \textit{\textbf{L2PGD}}\\ \hline
\multirow{2}{*}{MNIST}            
& Lenet 5 & 99.92$\%$ & 99.95$\%$  & 79.56$\%$ & 4.16$\%$ (90.84$\%$)   & 99.04$\%$ \\
& M. Keras CNN & 99.76$\%$ & 99.65$\%$  & 59.89$\%$ & 4.75$\%$ (88.99$\%$)  & 99.58$\%$ \\ \cline{1-7} 
\multirow{1}{*}{CIFAR-10}          
& C. Keras CNN & 100$\%$ & 100$\%$  & 96.17$\%$ & 98.39$\%$ (100$\%$)  & 100$\%$ \\ \bottomrule
\end{tabular}
\label{attack_success_before}
\end{table*}

\begin{table*}[h!]
\centering
\caption{\sc{Attack success rates when attacking the neural networks after entropic retraining}}
\begin{tabular}{lllllll}
\toprule
\multirow{2}{*}{\textbf{Data Set}} & \multirow{2}{*}{\textbf{Model}} & \multicolumn{5}{l}{\textbf{Attack Success Rates using the following methods:}}                                  \\ \cline{3-7} 
                                  &                                        & \textit{\textbf{LinfPGD}} & \textit{\textbf{BIM}} &  \textit{\textbf{FGSM}} & \textit{\textbf{C\&W}}\footref{CW_footnote} &  \textit{\textbf{L2PGD}}\\ \hline
\multirow{2}{*}{MNIST}            
& Lenet 5 & 14.48$\%$ & 3.48$\%$  & 2.95$\%$ & 0.58$\%$ (90.58$\%$)  & 5.19$\%$ \\
& M. Keras CNN & 1.07$\%$ & 0.06$\%$  & 0.06$\%$ & 0.01$\%$ (91.16$\%$)  & 0.72$\%$ \\ \cline{1-7} 
\multirow{1}{*}{CIFAR-10}          
& C. Keras CNN & 83.15$\%$ & 80.05$\%$  & 75.75$\%$ & 67.03$\%$ (94.90$\%$)  & 69.41$\%$ \\ \bottomrule
\end{tabular}
\label{attack_success_after}
\end{table*}

\subsection{Baselines}\label{sec:baselines}
\begin{table}[]
\centering
\caption{\sc{Accuracy of the original neural networks}}
\begin{tabular}{lllllll}
\toprule
\multirow{2}{*}{\textbf{Data set}} & \multirow{2}{*}{\textbf{Model}} & \multicolumn{5}{p{3.8cm}}{\textbf{Tested with:}}                                  \\ \cline{3-7} 
                                  &                                        & \textit{\textbf{Original}} & \textit{\textbf{$l_{2}$}\textbf{Noise}} & \textit{\textbf{$l_{\infty}$\textbf{Noise}}}\\ \hline
\multirow{2}{*}{MNIST}           
& Lenet 5   & 99.05$\%$ & 98.59$\%$  & 98.51$\%$             \\
& M. Keras CNN & 99.13$\%$ & 98.51$\%$  & 98.46$\%$            \\ \cline{1-7} 
\multirow{1}{*}{CIFAR-10}          
& C. Keras CNN  & 83.10$\%$ & 80.80$\%$  & 80.90$\%$  \\ \bottomrule
\end{tabular}
\label{accuracy_before}
\end{table}

To assess the impacts of entropic retraining and the characteristics of the resulting robust models, we defined the following baselines.
We used the pretrained models as introduced in \Cref{target_models} and performed two experiments.
First, we observed the models' performance classifying the benign original data, as well as noisy input images.
With this experiment we evaluated whether entropic retraining increases the sensitivity of the MUTs towards slight but unintentional perturbations.
To create the noisy input images we again used the \textit{foolbox} framework and generated two sets of test images.
For both sets, \textit{foolbox} generated noise which is added to the original input images.
Here, we chose the noise to either be $l_{2}$ or $l_{\infty}$ constrained.
This allows a direct comparison to the performance of the NNs when classifying adversarial examples bound by the same distance metrics.
Hence, in the MNIST setting, for the $l_{2}$ and $l_{\infty}$ constrained noise generation we set $\epsilon$ to $4.5$ and $0.3$, respectively.
Similarly, in the CIFAR-10 setting we chose $\epsilon$ to be $0.95$ and $0.03$.
In \Cref{accuracy_before} we summarize the resulting accuracy values.
For the two data sets and three MUTs we observe that noisy images are classified with a similar accuracy compared to their original counterparts.
The impact of noise on the performance of all MUTs is negligible as it decreases the accuracy scores by less than $5\%$.
We conclude, that the performance of the originally pretrained MUTs is robust towards noise.

Secondly, we evaluated the robustness of the pretrained MUTs by performing attacks with the five introduced methods.
As shown in \Cref{sec:how_robust}, we quantify the NNs' robustness by evaluating the attack success rates using $l_{2}$ and $l_{\infty}$ constrained attacks.
Note, that we solely used images which are originally classified correctly by the MUTs to measure the attack success rates.
In \Cref{attack_success_before}, we summarize the achieved rates.
We clearly see, as also seen in previous work, unsecured models are easily attacked.
For the three MUTs, the attacks achieved nearly perfect scores and show the low level of robustness in this case.

\section{Evaluation} \label{Results}
The Tables \ref{attack_success_after} and \ref{accuracy_after} summarize the main results of our paper for the three models after entropic retraining.
Again, we evaluated the performance against original and noisy images, as well as the attack success rates.
In both cases, we did not change any parameters unrelated to entropic retraining compared to the experiments using the original models shown in \Cref{sec:baselines}.

\begin{table}[]
\centering
\caption{\sc{Accuracy of the neural networks after entropic retraining}}
\begin{tabular}{lllllll}
\toprule
\multirow{2}{*}{\textbf{Data set}} & \multirow{2}{*}{\textbf{Model}} & \multicolumn{5}{p{3.8cm}}{\textbf{Tested with:}}                                  \\ \cline{3-7} 
                                  &                                        & \textit{\textbf{Original}} & \textit{\textbf{$l_{2}$}\textbf{Noise}} & \textit{\textbf{$l_{\infty}$\textbf{Noise}}}\\ \hline
\multirow{2}{*}{MNIST}           
& Lenet 5   & 97.73$\%$ & 95.19$\%$  & 94.90$\%$        \\
& M. Keras CNN & 99.07$\%$ & 98.28$\%$  & 98.16$\%$        \\ \cline{1-7} 
\multirow{1}{*}{CIFAR-10}          
& C. Keras CNN   & 85.60$\%$   & 83.70$\%$  & 83.70$\%$ \\ \bottomrule
\end{tabular}
\label{accuracy_after}
\end{table}

\Cref{accuracy_after} shows the accuracy scores of the three MUTs when classifying the original as well as noisy inputs.
For MNIST, Lenet's performance decreased by $1.3\%$ when classifying the original inputs.
For the noisy inputs, the accuracy scores decreased by $3.5\%$ and $3.7\%$ for the $l_{2}$ and $l_{\infty}$ constrained cases, respectively.
The capability of the Keras example model classifying MNIST images decreased by $0.1\%$.
In this case, the $l_{2}$ and $l_{\infty}$ constrained noisy images were classified $0.2\%$ and $0.3\%$ less accurate.
Interestingly, for the CIFAR-10 MUT we find an increase in accuracy of $3\%$ when classifying the original inputs. 
This is also visible for noisy images, here the model performed $3.59\%$ and $3.46\%$ better for the $l_{2}$ and $l_{\infty}$ constrained examples.
In summary, we conclude that entropic retraining does not decrease the natural performance of the models and in some cases even provides higher accuracy values.
Furthermore, the robustness against noise did not decrease and the MUTs achieved similar levels of accuracy compared to their initial versions.

\footnotetext[1]{\label{CW_footnote}In some cases, the C$\&$W attack produced black-only images. We discarded these for the calculation of the attack success rate. The rate in the parentheses is produced including the black-only images.}

More importantly, we want to focus on the resulting levels of robustness for the three MUTs.
Thus, in \Cref{attack_success_after} we present the main results of our paper.
We observe a significant increase in robustness for all data sets, MUTs, and attack methods.
For the MNIST data set, only one attack achieved a success rate greater than $10\%$.
This is the case when attacking Lenet 5 with the $l_{\infty}$ bound PGD method.
For the remaining scenarios, we measured success rates of well below $10\%$ and in some cases even close to $0\%$.
Evaluating the CIFAR-10 data set we see a significant decrease in the attack success rates as well.
In the worst case, for the $l_{\infty}$ PGD attack, the attack success rate decreased by $16.85\%$.
For the $l_{2}$ bound PGD and C\&W methods the success rates decreased by more than $30\%$.

In summary, we find increased levels of robustness without loosing the capability of reliably classifying original as well as noisy inputs.

\section{Discussion} \label{Discussion}
Our adapted loss function allows an information-theoretic lossy but robust training process for NNs.
We observe that entropic retraining generalizes well and achieves good results for two data sets and different NN architectures.
Note, for the sake of simplicity we kept the parameter optimization in our experiments as short as possible.
We used similar settings for all three MUTs to show the simplicity of our approach.
A deliberate choice of parameters with respect to the chosen architecture and use-case may further improve the results.
For example, the number of epochs heavily influences the trade-off between robustness and accuracy.
Future work may evaluate the performance of the approach using more complex data sets and NN architectures as well as an extensive hyperparameter search to understand the impacts of the settings to their full extent.
Our positive results clearly show the need for more research leveraging these initial observations.
We expect that more work in this direction will lead to an even clearer picture and understanding of the meaning of \textit{robustness}.
To this end, we hope that our findings help to approach the difficult question of how to reliably quantify and improve NN robustness.

\section{Conclusion} \label{Conclusion}
In this paper we present, implement, and evaluate a new training approach for NNs we call entropic retraining.
Based on our information-theoretic-inspired analysis of adversarial training we gain new insights in the information flow of robust models.
This motivates us to adapt and expand currently used loss functions. 
With entropic retraining, we optimize for accuracy and robustness simultaneously.
Our evaluation clearly indicates the effectiveness of entropic retraining for multiple data sets and NN architectures.

\bibliographystyle{IEEEtranS}
\bibliography{mybib}{}

\end{document}